\documentstyle[12pt]{article}
\begin{document}

\begin{center}  {\Large {\bf Pedestal and Peak Structure in
Jet Correlation}}
\vskip .75cm
   {\bf Charles B.\ Chiu$^1$ and  Rudolph C. Hwa$^{2}$}
\vskip.5cm

   {$^1$Center for Particle Physics and Department of Physics\\
University of Texas at Austin, Austin, TX 78712, USA\\
\bigskip
$^2$Institute of Theoretical Science and Department of
Physics\\ University of Oregon, Eugene, OR 97403-5203, USA}
\end{center}

\begin{abstract}
We study the characteristics of correlation between particles
in jets produced in heavy-ion collisions. In the framework of
parton recombination we calculate the $\eta$ and $\phi$
distributions of a pion associated with a trigger particle. The
origin of the pedestal in $\Delta\eta$ is related to the
longitudinal expansion of the thermal partons that are
enhanced by the energy loss of hard partons traversing the
bulk medium. The peaks in $\Delta\eta$ and $\Delta\phi$ are
related to the same angular spread of the shower partons in a
jet cone. No artificial short- or long-range correlations are
put in by hand. A large part of the correlation between hadrons in jets is
due to the correlation among the shower partons arising from
momentum conservation. Recombination between thermal and
shower partons dominates the correlation characterisitics in
the intermediate $p_T$ region.

\vskip0.5cm
PACS numbers:  25.75.Gz, 25.75.-q
\end{abstract}
\section{Introduction}

The discovery of the medium effects on jets produced at the relativistic
heavy-ion collider (RHIC) has contributed greatly to the understanding
of the physics underlying hard partons traversing a hot and dense
quark-gluon system\cite{fr,pj}.  Subsequently, a large number of
experimental investigations have revealed details of the properties of
jets, both on the near side and on the away side, in heavy-ion collisions
(HIC) \cite{jaw}-\cite{jat}.  Particle correlations within jets are shown to
have distinctive properties in the azimuthal angles $\phi$ and in
pseudorapidities $\eta$ of the associated particles.  Of particular
noteworthiness is the peak in $\Delta \eta$ between the $\eta$ values
of the trigger and the associated particles on the near side:  it sits above
a flat plateau or a pedestal \cite{jaw}.   No pedestal is found in $\Delta \phi$,
although a peak in that variable is prominent.  It is our aim in this paper
to study the peaks and pedestal in $\Delta \phi$ and $\Delta \eta$ and
to discuss the physical origins of these jet characteristics.

Our approach to the problem of correlation between particles in jets will
be based on parton recombination that has been shown to be successful
in reproducing the single-particle distributions in Au+Au collisions
\cite{hy}, and in d+Au collisions \cite{hy2}.  It has also been applied to
the study of dihadron correlations \cite{hy3,ht}.  However, since those
investigations were carried out in the framework of the one-dimensional
(1D) formulation of the recombination model, a generalization to
account for the 3D geometry of the problem is necessary if we are to
describe the features
in the $\phi$ and $\eta$  variables.  Once we consider the 3D aspects of
jets the angular distribution of the shower partons in the jet cone
becomes an important new feature. Our treatment will be
phenomenological, since no results from perturbative QCD
considerations can reliably be applied to the intermediate $p_T$ region,
which is where the data on correlations are taken.  Furthermore, the
pedestal phenomenon suggests that the jet interacts with the
environment in a way that is sensitive to the hydrodynamical expansion
of the medium, since there is longitudinal expansion, but not azimuthal
expansion.  It turns out that we can relate the pedestal to the local
enhancement of the thermal bath due to the energy loss of the hard
parton to the medium.

Our approach to the problem, though not based on first principles, provides
the first theoretical interpretation of the pedestal and peak structure of
the jets produced at RHIC in quantitative terms. The roles that shower and
thermal partons play in the hadronization process are found to be crucial
and cannot be reinterpreted in any other model that we are aware of.

In Sec.\ 2 we describe how the 3D properties of jets are incorporated in the
   recombination formalism. In Sec.\ 3 the pedestal phenomenon is
investigated, followed by the study of the peaks in $\Delta\eta$ and
$\Delta\phi$ distributions in Sec.\ 4. The conclusion is given in the last
section.

\section{Two-Particle Distribution}

In the framework of parton recombination formulated in 1D for
particle production in HIC \cite{hy}-\cite{ht}, let us start by writing down
the single-and two-particle distributions
\begin{eqnarray}
{dN_{\pi_{1}}  \over  p_1 dp_1} = { 1 \over p_1^2 }\int {dq_1 \over
q_1}{dq_2 \over  q_2}F_2 (q_1, q_2) R_{\pi_{1}}(q_1, q_2, p_1)
\quad
\label{1}
\end{eqnarray}
\begin{eqnarray}
{dN_{\pi_1 \pi_2}  \over  p_1 p_2 dp_1d p_2} =  {1 \over
p^2_1 p^2_2}
\int\left(\prod^4_{j=1}{dq_j  \over  q_j}\right) F_4
(q_1, q_2,q_3, q_4)  R_{\pi_1}(q_1, q_3, p_1)
R_{\pi_2}(q_2, q_4, p_2) \ ,
\label{2}
\end{eqnarray}
where the recombination function (RF) for a pion is \cite{hy4}
\begin{eqnarray}
R_{\pi}(q_1, q_2, p) ={q_1q_2  \over p^2} \delta\left({q_1
\over p}+ {q_2  \over p} - 1\right)\ .
\label{3}
\end{eqnarray}
The two- and four-parton distributions, $F_2$ and $F_4$, can be
written in terms of their components as
\begin{eqnarray}
F_2(1,2) = ({\cal TT} + {\cal TS}  + {\cal SS})_{12}\ ,
\label{4}
\end{eqnarray}
\begin{eqnarray}
F_4(1,2,3,4) = ({\cal TT} + {\cal TS} + {\cal SS}) _{13} ({\cal TT} + {\cal
TS} + {\cal SS}) _{24} \ .
\label{5}
\end{eqnarray}
The thermal parton distribution is
\begin{eqnarray}
{\cal T}(q) =
Cqe^{-q/T} \ ,
\label{6}
\end{eqnarray}
and the shower partons in a jet have the form
\begin{eqnarray}
{\cal S}(q) = \xi \sum_i \int dk k
f_i(k) S^j_i (q/k) \ ,
\label{7}
\end{eqnarray}
where $f_i(k)$ is the distribution of hard parton $i$ in a HIC, and $\xi$
is the average fraction of hard partons that emerge from the thermal
medium to hadronize.  $S^j_i$ is the shower-parton distribution (SPD)
of parton $j$  in a shower initiated by a hard parton $i$; their
parameterizations for various $i$ and $j$ are given in \cite{hy5}. 
For the sake of clarity we have omitted the labels for parton species 
for $\cal T$ and $\cal S$ in the above equations and in what follows, 
except when the identity of a parton is needed for emphasis.

It is important to recognize that Eq.\ (\ref{7}) should not be applied to
every ${\cal S}$ that appears in Eqs.\ (\ref{4}) and (\ref{5}) because all
the shower partons in $F_2$ and $F_4$ are created in the same jet, so
all the ${\cal SS}\cdots$ terms share the same hard parton with
momentum $k$. Thus for two shower partons $({\cal SS})$, their joint
distribution is
\begin{eqnarray}
({\cal SS})(q_1, q_2)=\xi\sum_i
\int dk k f_i(k) \left\{S^j_i\left({q_1\over
k}\right),S^{j'}_i\left({q_2\over k-q_1}\right)
\right\},
\label{8}
\end{eqnarray}
where the quantity in the curly brackets has the form in terms of
momentum fractions $x_1$ and $x_2$
\begin{eqnarray}
\left\{S^j_i (x_1),\ S^{j'}_i \left({x_2\over 1-x_1}\right)\right\} =
{1\over  2} \left[S^j_i (x_1) S^{j'}_i \left({x_2\over 1-x_1}\right) +
S^j_i
\left({x_1\over 1-x_2}\right)S^{j'}_i (x_2)\right] .
\label{9}
\end{eqnarray}
This is done to guarantee momentum conservation $x_1 + x_2 \leq 1$
and to symmetrize the order of emission of the two partons.  Not only
should Eq.\ (\ref{8})  apply to two adjacent shower partons, like in $({\cal
SS})_{13}$ and $({\cal SS})_{24}$, which recombine to form the pions at
$p_1$ and $p_2$, respectively, but also to two shower partons like in
$({\cal TS})_{13}({\cal TS})_{24}$, which do not recombine with
each other, but separately with thermal partons.  It is the structure in
Eq.\ (\ref{8}) that gives rise to correlation between two shower partons
in a jet, and thereby endows the detected hadrons at high $p_T$ with
correlation. However, that is sufficient primarily in the case of
considering only the correlation among momentum magnitudes in 1D.

To generalize our consideration to 3D, let us focus our attention first on
the $({\cal TS})$ components in Eq.\ (\ref{5}) that give the most
important contribution to the trigger and associated particles in central
collision.  To be specific let us further designate the trigger particle to be
a $\pi^+$ and the associated particle to be a $\pi^-$.  The 3D
expression for Eq.\ (\ref{2}) in that case is
\begin{eqnarray}
{dN_{\pi^+ \pi^-}  \over  p_1 dp_1 d\eta_{tr}d\phi _{tr} p_2
dp_2 d\eta d\phi} &=&   {1 \over (p_1 p_2)^2}
\xi \sum_i \int dk kf_i(k)\int\left(\prod^4_{j=1}d^3r_j{d^3q_j  \over
q_j}\right)\nonumber\\
&&\times S^u_i\left(\vec{r}_1, \vec{q}_1\right) {\cal
T}^{\bar{d}}\left(\vec{r}_3,
\vec{q}_3\right)
R^{(3)}_{\pi^+}\left(\vec{r}_1, \vec{r}_3, \vec{r}_{p_1};
\vec{q}_1,
\vec{q}_3,\vec{p}_1\right) \nonumber\\
&&\times S^d_i\left(\vec{r}_2, \vec{q}_2\right) {\cal
T}^{\bar{u}}\left(\vec{r}_4,
\vec{q}_4\right)
R^{(3)}_{\pi^-}\left(\vec{r}_2, \vec{r}_4, \vec{r}_{p_2}; \vec{q}_2,
\vec{q}_4,\vec{p}_2\right)
\label{10}
\end{eqnarray}
where $R^{(3)}$ is the 3D version of the RF and it includes the spatial
coordinates.  Since recombining partons must not only have collinear
momenta but also have overlapping wave functions, there are many
narrow Gaussian distributions in $R^{(3)}$ that reduce the phase space
of integration.  Let us first restrict our attention to the $x$-$z$
plane, i.e.\
the plane containing the hard parton momentum $\vec{k}$ and the
longitudinal direction $\hat{z}$, and assume that all $\vec{q}_i$
vectors are in that plane so that we may write
\begin{eqnarray}
\vec{q}_i = (q_i, \theta_i),\quad \vec{p}_1 = (p_1,
\theta_{trig}),\quad \vec{p}_2 =
(p_2,
\theta)
\label{11}
\end{eqnarray}
and
\begin{eqnarray}
\eta_i = \ln \, \cot {\theta_i \over 2},\quad \eta_{trig} = \ln \, \cot
{\theta_{trig}
\over 2},\quad \eta = \ln \, \cot {\theta \over 2} .
\label{12}
\end{eqnarray}
We further assume that $\vec{q}_1$ is in the direction of $\vec{k}$ and
that the trigger momentum $\vec{p}_1$ is also along $\vec{k}$.  This is
a simplification that does not compromise the angular correlation
between the trigger and associated particles, which we shall study in
detail.  Since $R^{(3)}$ requires the wave functions of the recombining
partons to overlap, the shower and thermal partons at $\vec r_1$ and $\vec
r_3$ should be nearby. These assumptions allow us to reduce the trigger
momentum part of Eq.\ (\ref{10}) to the usual 1D formulation
\begin{eqnarray}
\int d^3r_1d^3r_3 {d^3q_1  \over q_1 } {d^3q_3  \over q_3 }
S^u_i\left(\vec{r}_1, \vec{q}_1\right) {\cal
T}^{\bar{d}}\left(\vec{r}_3,
\vec{q}_3\right)
R^{(3)}_{\pi^+}\left(\vec{r}_1, \vec{r}_3, \vec{r}_{p_1};
\vec{q}_1,
\vec{q}_3,\vec{p}_1\right) \nonumber\\
=\left. \int {dq_1  \over q_1 } {dq_3  \over q_3 }
S^u_i\left({q_1 \over  k}\right) {\cal T}\left(q_3 ,
\theta_3\right)\right|_{\theta _3=\theta _1=\theta _{trig}}
R^{(1)}_{\pi^+}\left(q_1, q_3, p_1\right) ,
\label{13}
\end{eqnarray}
where $R^{(1)}$ is the usual RF in 1D, given in Eq.\ (\ref{3}).  The 
superscript $\bar d$ on $\cal T$ is omitted on the RHS since the 
flavor dependence of the thermal partons is negligible and ignored. 
We shall
assume that at midrapidity, $- 0.7 < \eta_{trig}< + 0.7$, the thermal
distribution does not contain essential dependence on $\eta_3$ so that
the condition $\theta _3=\theta _1=\theta _{trig}$ does not lead to any
restriction that would cause $ {\cal T}\left(q_3 , \theta_3\right)$ to
deviate from the usual parameterization given in Eq.\ (\ref{6}).

In the case of the associated particle we must recognize that the shower
parton $\vec{q}_2$ may be emitted at an angle $\psi$ relative to the
hard parton momentum $\vec{k}$.  Since $\vec{k}$ forms an angle
$\theta _1$ with the $z$ axis (after identifying the direction $\hat{q}_1$
with
$\hat{k}$), we have
\begin{eqnarray}
\psi = \theta _2 - \theta _1 .
\label{14}
\end{eqnarray}
The recombination of the shower parton at $\vec{q}_2$ with a thermal
parton at $\vec{q}_4$ requires that they are not only collinear, but also
overlapping spatially.  There are therefore $\delta$-functions,
$\delta(\theta _2 - \theta _4) \delta \left({ \theta _2 + \theta _4 \over
2} - \theta\right)$, that force the momentum $\vec{p}_2$ of the
associated particle to be in the same direction as $\hat{q}_2$ and
$\hat{q}_4$.  The corresponding 3D integrations result in
\begin{eqnarray}
\int d^3r_2d^3r_4 {d^3q_2  \over q_2 } {d^3q_4  \over q_4 }
S^d_i\left(\vec{r}_2, \vec{q}_2\right) {\cal
T}^{\bar{u}}\left(\vec{r}_4,
\vec{q}_4\right)
R^{(3)}_{\pi^-}\left(\vec{r}_2, \vec{r}_4, \vec{r}_{p_2};
\vec{q}_2,
\vec{q}_4,\vec{p}_2\right) \nonumber\\
=\left. \int {dq_2  \over q_2 } {dq_4  \over q_4 }
S^d_i\left({ q_2 \over  k-q_1},\psi\right) {\cal T}\left(q_4 ,
\theta_4\right)\right|_{\theta _4=\theta_2 =\theta _1+\psi}
R^{(1)}_{\pi^-}\left(q_2, q_4, p_2\right) \ .
\label{15}
\end{eqnarray}
This equation is very similar to Eq.\ (\ref{13}) but with one crucial
difference; that is, the shower parton distribution now depends on
$\psi$.  In Eq.\ (\ref{13}) no angular dependence of the shower parton
$\vec{q}_1$ is assumed for simplicity, since it is only the difference
between $\vec{q}_1$ and $\vec{q}_2$ that matters.  That angular
difference, $\psi$, made explicit in Eq.\ (\ref{14}), is translated into the
angular difference between $\vec{p}_1$ and $\vec{p}_2$.  We have
tacitly assumed that the thermal partons have local angular spread that
allows a $\vec{q}_3$ to recombine with $\vec{q}_1$ to
form $\vec{p}_1$, and similarly $\vec{q}_4$ with $\vec{q}_2$ to form
$\vec{p}_2$, in such a way that $\vec p_1$ and $\vec p_2$ are in the 
directions of $\vec q_1$ and $\vec q_2$, respectively.

The dependence of a SPD on $\psi$ cannot be determined from pQCD,
since the values of $p_T$ concerned is in the $1 < p_T <4$ GeV/c range.
We shall assume that the SPD in Eq.\ (\ref{15}) can be written in the
following factorizable form
\begin{eqnarray}
S^d_i\left({q_2 \over k-q_1 },\psi\right) = S^d_i\left({q_2 \over k-q_1
}\right) G \left(\psi, {q_2 \over k} \right) ,
\label{16}
\end{eqnarray}
where the dependence on $\psi$ has a Gaussian distribution
\begin{eqnarray}
G(\psi, x)= exp \left[ - {\psi^2  \over  2 \sigma ^2(x)}\right] \ ,
\label{17}
\end{eqnarray}
whose half-width depends on the momentum fraction $x$ of the shower
parton
\begin{eqnarray}
\sigma (x) = \sigma _0 (1 - x)
\label{18}
\end{eqnarray}
with $\sigma_0$ being a parameter that is adjustable.  $S^d_i$ on the 
RHS of Eq.\ (\ref{16}) is to be symmetrized with $S^u_i$ in Eq.\
(\ref{13}) without further complication from angular consideration.
The momentum fraction relevant for $\sigma(x)$ is $q_2/k$,
independent of $q_1$.  Once we obtain Eq.\ (\ref{16}) where $G(\psi,
q_2/k)$ stands as a modifying factor, it is reasonable to liberate that
factor from the way in which it is derived, and allow it to assume a 3D
property so that when we later consider the $\Delta\phi$ behavior in
the transverse plane the same factor $G(\psi, q_2/k)$ applies to the
azimuthal angle between $\vec{p}_1$ and $\vec{p}_2$ with $\psi$ replaced by
$\Delta\phi$. Using Eqs.\ (\ref{12}) and (\ref{14}) we obtain with $\theta =
\theta_2$
\begin{eqnarray}
\tan {\psi  \over  2} =\ tan {\theta-\theta_1  \over  2} = g(\eta,\eta_1) =
{e^{-\eta} -e^{-\eta_1}
\over  1+ e^{-\eta-\eta_1}} \quad ,
\label{19}
\end{eqnarray}
where $\eta_1$ is to be identified with $\eta_{trig}$.

We now can write Eq.\ (\ref{10}) with the help of Eqs.\ (\ref{3}),
(\ref{13}) and (\ref{15}) in the simpler form
\begin{eqnarray}
{dN^{TSTS}_{\pi^+ \pi^-}  \over  p_1 dp_1 d\eta_1d\phi _1 p_2
dp_2 d\eta d\phi} =   {1 \over (p_1 p_2)^3}
\xi \sum_i \int dk kf_i(k)\int dq_1\int dq_2 \theta (k -
q_1-q_2)\nonumber\\
\sum_{j,j^{\prime}} {\cal T}(p_1-q_1,\eta_1)\left\{ S^j_i \left({q_1  \over  k}
\right), S^{j^{\prime}}_i \left({q_2  \over  k-q_1}\right)\right\} {\cal
T}\left(p_2-q_2, \eta\right)\nonumber\\ \left. G\left(\psi, {q_2  \over  k}
\right)\right|_{\psi = 2 tan ^{-1}g(\eta, \eta_1)}
\label{20}
\end{eqnarray}
where $i$ is summed over all hard parton species. If $i$ is a valence quark
of $\pi^+$ or $\pi^-$, the corresponding SPD is $K$, while if $i$ is 
a sea quark, the
corresponding SPD is $L$ in the notation of \cite{hy5}, where 
$K=S_i^{val+sea}$ and $L=S_i^{sea}$.  The integral over $k$ will, in
practice, be from 3 to 30 GeV/c, and $q_i$ will be integrated from $0$ to
$p_i$, although the formalism is not reliable for any transverse momentum
less than 1 GeV/c.  There are other terms of $F_4$ contained in Eq.\
(\ref{5}) that can be written out as in Eq.\ (\ref{20}), but will not be
detailed here.

\section{Pedestal and Background Subtraction}

The experimental procedure of making background subtraction involves
multiple considerations, and the result is that there remains a residual
pedestal in $\Delta \eta$ in central collisions, but no pedestal in $\Delta
\phi$, where $\Delta \eta$ and $\Delta \phi$ are differences in $\eta$
and $\phi$ between the associated particle and the trigger.  In
\cite{hy3} the dihadron correlation is calculated with the assumption
that the factorizable part of the two-particle distribution corresponds to
the background.  That leaves only the $({\cal TS + SS})_{13} ({\cal
TS + SS})_{24}$ terms of $F_4$ to consider.  We now find that in order
to understand the origin of the pedestal what constitutes the
background must be reconsidered.

The first point that we want to address is the effect of energy loss of
hard partons traversing the hot medium.  The parameter $\xi$ in  Eqs.\
(\ref{7}) and (\ref{8}), being only 0.07 as determined in \cite{hy} for
central Au+Au collisions, implies that a large fraction of the hard
partons produced are absorbed or attenuated by the bulk medium.
Those that emerge to hadronize outside must have been created near
the surface.  In the short distance that such a parton travels in the
medium it must on average lose some energy and locally enhance the
thermal motion of the partons in the environment.  Those enhanced
thermal partons should have an effective inverse slope $T$ that is
slightly higher (by $\Delta T$) than that of the usual thermal partons not
influenced by the passage of any hard partons.  The latter corresponds
to what is measured for the low-$p_T$ pions in single-particle
distribution, which for $p_T < 2$ GeV/c has led to the determination of
$T = 0.317$ GeV/c in \cite{hy}.  If $\Delta T$ is small, as we expect it to
be, it does not make too much difference in the calculation of the
single-particle distribution in the intermediate $p_T$ region.  However,
in principle, in thermal-shower recombination the shower partons
recombine with the enhanced thermal partons that are in the immediate
vicinity of the hard parton that creates the shower.  To distinguish the
two values of the inverse slope, let us use $T$ to denote the one for the
enhanced thermal parton, since that is the only physically relevant one
for ${\cal TS}$ recombination, i.e., it is what appears in Eq.\ (\ref{6})
for ${\cal T}$.  Let us use $T_0$ to denote the inverse slope for the
thermal medium undisturbed by jet quenching, i.e., $T_0 = 0.317$ GeV/c.
We define
\begin{eqnarray}
\Delta T = T - T_0 \quad ,
\label{21}
\end{eqnarray}
which is a parameter to be determined phenomenologically in this
paper, although it can presumably be determined directly by dedicated
experimentation.  We shall relate $\Delta T$ to the pedestal.  The
physics involved is sensible, since the pedestal is seen only for central
collisions where ${\cal TS}$ recombination is important.  In peripheral
collisions thermal partons play a minor role and there are few of them to be
enhanced.

We now give an argument why the pedestal is seen in $\Delta \eta$, but
not in $\Delta \phi$.  When a hard parton is first scattered or created at
a large angle relative to the incident beam direction, the bulk medium is
in a highly compressed state.  As it expands longitudinally, the region of
enhanced thermal partons expands with the whole system, although
limited to the neighborhood of where the hard parton traverses the thin
layer close to the surface.  Thus a section at midrapidity can have the
enhanced $T$.  In the transverse plane, on the other hand, the
expansion is in the radial direction only; that is, there is no mixing
across different azimuthal sections.  The subtraction scheme carried out
by STAR \cite{jaw} defines the yield $dN/d\Delta \phi$ to be zero at
$|\Delta \phi| = 1$ after subtraction.  Since there is no expansion in the
$\phi$ direction, we can take this subtraction scheme to imply that all
thermal partons that are enhanced stay inside the $|\Delta \phi| < 1$
region, and that for $|\Delta \phi| > 1$ there is no thermal-thermal
recombination that is not in the background.  There may still be an
enhancement of the yield inside the $|\Delta \phi| < 1$
region that plays a role similar to the
pedestal in $\Delta \eta$, but it cannot extend beyond $|\Delta \phi| =
1$ by virtue of the subtraction scheme.

The above description cannot be demonstrated by a transport model,
since we do not have a Monte Carlo code for the evolution process.
However, the physical content of our discussion will be embodied in our
quantitative formulation below.  The recombination model actually
provides a more transparent description of the hadronization process
than what a code without reliable equations can offer.

Let us first consider the particles that are associated with a trigger.
Since in Au+Au collisions the ${\cal TS}$ component is dominant in
the intermediate $p_T$ region \cite{hy}, which includes the trigger
window $4 < p_T < 6$ GeV/c, we select the $({\cal TS})_{13}$ term in
Eq.\ (\ref{5}) for our discussion below, although both terms in $({\cal
TS+SS})_{13}$ are included in our calculation.  Thus the parts of $F_4$ that
contribute to the trigger and its associated particles are
\begin{eqnarray}
F^{tr + as}_4 = ({\cal TS})_{13} ({\cal TT} + {\cal TS} + {\cal SS})_{24}
\quad .
\label{22}
\end{eqnarray}
Among the three terms on the right side it is clear that $({\cal TS} +
{\cal SS})_{24}$ are directly related to the jet and contribute to the peak
in the associated particle distribution (APD), while $({\cal TT})_{24}$
does not involve the shower partons but can nevertheless contribute to
the APD outside the peak.  It is therefore our prime candidate for the
pedestal.  The absence of an obvious pedestal in the APD in $\Delta
\phi$ is a consequence of the background subtraction scheme in the
analysis of the experimental data, where the $\Delta \phi$ distribution
at $|\Delta \phi| = 1$ is identified to be the background.  In our
description of the APD the background corresponds to
\begin{eqnarray}
F^{bg}_4 = ({\cal TS})_{13} ({\cal T}_0 {\cal T}_0)_{24} \quad ,
\label{23}
\end{eqnarray}
where ${\cal T}_0(q)$ is the thermal parton distribution in the absence
of any hard partons, i.e.,
\begin{eqnarray}
{\cal T}_0 (q) = Cqe^{-q/T_0} \quad .
\label{24}
\end{eqnarray}
The meaning of $T_0$ has already been discussed in connection
with Eq.\ (\ref{21}).  We have tacitly assumed that the normalization
factor $C$ is unchanged from that in Eq.\ (\ref{6}).  The experimental
subtraction scheme implies that
\begin{eqnarray}
({\cal TS})_{13} ({\cal TT})_{24} = ({\cal TS})_{13} ({\cal T}_0 {\cal
T}_0)_{24}
\label{25}
\end{eqnarray}
at $|\Delta \phi| = 1$.  We now define the APD with background
subtracted to be generated by
   \begin{eqnarray}
F^{AP}_4 = ({\cal TS})_{13} ({\cal TT}-{\cal T}_0 {\cal T}_0+{\cal
TS}+{\cal SS})_{24}
\quad ,
\label{26}
\end{eqnarray}
in which $({\cal TT}-{\cal T}_0 {\cal T}_0)_{24}$ vanishes at  $|\Delta
\phi| = 1$, but need not be zero at  $|\Delta \phi| < 1$.  Equation
(\ref{26}) is our basic input in the calculation of the APD.

Let us now write our two-particle distribution in a form suitable for
application of the experimental cuts.  In Eq.\ (\ref{20}) we have a
differential distribution in $dp_1d\eta_1d\phi_1dp_2d\eta d\phi$
where $\eta_1$ and $\phi_1$ denote the variables of the trigger, and
$\eta$ and $\phi$ those of the associated particle.  With the definition
\begin{eqnarray}
\Delta \eta = \eta - \eta_1 , \quad \Delta \phi = \phi - \phi_1 \quad ,
\label{27}
\end{eqnarray}
we change the differentiated variables to
$dp_1 dp_2 d\eta_1d\Delta\eta d\Delta\phi$, with $\phi_1$ being used
as a free reference point in the assemblage of the $\Delta\phi$
distribution; the trigger $\eta_1$ is to be integrated over the trigger
window $- 0.7 < \eta_1 < +0.7$.  We then have
\begin{eqnarray}
N(1,2)&\equiv& {dN_{\pi_1 \pi_2}  \over  dp_1 dp_2 d\eta_1d \Delta
\eta d \Delta\phi}  \nonumber\\  &= &{1
\over p_1 p_2}
\int\left(\prod^4_{j=1}{dq_j \over  q_j}\right) F^{AP}_4
(q_1, q_2, q_3, q_4)  R_{\pi_1}(q_1, q_3, p_1)
R_{\pi_2}(q_2, q_4, p_2) \ .
\label{28}
\end{eqnarray}
The experimental APD
involves integrations over the trigger window in $\eta_1$ and $p_1$ and
the window of the associated particle in $p_2$.  It is
\begin{eqnarray}
{ dN^{AP} \over  d\Delta \eta d \Delta \phi} = { \int^{p_b} _{p_a} dp_2 \int
^6_4 dp_1\int^{0.7} _{-0.7} d\eta_1 N(1,2)\over
\int ^6_4 dp_1
\int^{0.7} _{-0.7} d\eta_1 N(1) } \ ,
\label{29}
\end{eqnarray}
where $N(1)$ is the one-particle trigger distribution that receives
contribution from the $({\cal TS})_{13}$ component only, i.e.
\begin{eqnarray}
N(1)\equiv  {dN^{tr}_{\pi_1 }  \over  dp_1 d\eta_1}    = {1
\over p_1^2}
\xi \sum_i \int dk k f_i(k) dq_1 \sum _j {\cal T} (p_1-q_1, \eta_1)
S^j_i \left({q_1  \over k } \right)\quad.
\label{30}
\end{eqnarray}

The two-particle distribution $N(1, 2)$ in Eq.\ (\ref{29}) can be
divided into two explicit pieces.  For the remainder of this section we
consider only the piece related to the pedestal
\begin{eqnarray}
N(1,2)^{ped} &\equiv&  {dN^{ped}_{\pi_1 \pi_2}  \over  dp_1 dp_2 d\eta_1
d\Delta
\eta d \Delta \phi} \nonumber\\   &=& {1
\over (p_1 p_2)^2}
\xi \sum_i \int dk k f_i(k) dq_1 dq_2\nonumber\\
&& \sum _j  {\cal T} (p_1-q_1,
\eta_1) S^j_i \left({q_1  \over k } \right)\left[ {\cal
T}(q_2,\Delta \eta ) {\cal T}(p_2 - q_2,\Delta \eta )H (\Delta
\phi)\right.\nonumber\\
&&\left.- {\cal T}_0 (q_2){\cal T}_0 (p_2 - q_2) \right]\quad ,
\label{31}
\end{eqnarray}
where we have extracted the $\phi$ dependence of the associated
particle in the form of a factor $H (\Delta \phi)$, which satisfies some
constraints to be specified below.  It should be noted that Eq.\ (\ref{31})
is factorizable, since the two parts of $({\cal TS})_{13} ({\cal TT}-{\cal
T}_0 {\cal T}_0)_{24}$ are independent of each other.  The $({\cal
TS})_{13}$ part contributes to the trigger distribution given in Eq.\
(\ref{30}).  It will be cancelled in the ratio defined in Eq.\ (\ref{29}).

Using Eq.\ (\ref{24}), the integration
of the last term  in Eq.\ (\ref{31}) over $q_2$
can readily be carried out, giving
\begin{eqnarray}
{1  \over p^3_2 } \int^{p_2}_0 dq_2 {\cal T}_0 (q_2){\cal T}_0 (p_2 - q_2) = {
C^2 \over  6} \exp (-p_2/T_0) \quad ,
\label{32}
\end{eqnarray}
which, upon further integration over the window $(p_a,p_b)$ of the
associated particle, results in
\begin{eqnarray}
\int^{p_b}_{p_a}dp_2 p_2 {dN^{bg}  \over  p_2 dp_2 } = { 1 \over  6}(C
T_0)^2 h \left({p_a  \over T_0 }, {p_b  \over T_0 } \right) \quad
,
\label{33}
\end{eqnarray}
   where
\begin{eqnarray}
h(x, y)= (1 + x)e^{-x} - (1 + y)e^{-y} \quad .
\label{34}
\end{eqnarray}
Similar result follows for the ${\cal TT}$ term in Eq.\ (\ref{31}).

Next, we consider the $\Delta \eta$ and $\Delta \phi$ dependences of
the pedestal in Eq.\ (\ref{31}).  As we have discussed qualitatively in the
beginning of this section, the longitudinal expansion allows the enhanced
thermal partons to extend over a wide range in $\eta$, but radial
expansion does not increase the range in $\phi$.  Putting these
properties into quantitative terms, we give ${\cal T}(q_2)$ no essential
dependence on $\Delta \eta$ in Eq.\ (\ref{31}), but a Gaussian
dependence on $\Delta \phi$ so that
\begin{eqnarray}
H(\Delta\phi)= c \, \exp
\left(-\Delta\phi^{2}/2\sigma^2_{\phi}\right)\quad .
\label{35}
\end{eqnarray}
The normalization $c$ is to be determined by the condition that when
$\Delta\phi$ is integrated over the experimental window $(-0.5,+0.5)$
we obtain
\begin{eqnarray}
\int^{0.5}_{-0.5}d\Delta\phi H(\Delta\phi)= 1\quad .
\label{36}
\end{eqnarray}
In this way we can relate Eq.\ (\ref{31}) to the pedestal observed in the
experiment without explicit factors dependent on the experimental
windows.  The half-width $\sigma_{\phi}$ is adjusted so that the quantity
inside the square brackets in Eq.\ (\ref{31}) vanishes at $|\Delta
\phi| = 1$, as required by the subtraction scheme, Eq.\ (\ref{25}).  What
we do here is to fix all the extra free parameters of the problem by the
experimental features of the data.  Since our aim is to reproduce the
observed characteristics of the data, which are presented with specific
cuts, it is impossible to do so without incorporating those cuts.
However, it does not imply that we are merely fitting the data with free
parameters.  We shall perform several multi-dimensional integrals to
obtain in the next section the APD with the peak whose magnitude is a
prediction of our model calculation.

Upon integration of Eq.\ (\ref{29}) over $\Delta\phi$ in the acceptance
window, we get for the pedestal part
\begin{eqnarray}
{dN^{ped}  \over d\Delta\eta} = \int^{0,5}_{-0.5}d\Delta\phi {dN^{ped}
\over d\Delta\eta d\Delta\phi } = { C^2 \over  6} \left[T^2 h\left({p_a  \over
T } , {p_b  \over T } \right)- T^2_0 h\left({p_a  \over  T_0},
{p_b  \over  T_0 } \right)   \right] \ ,
\label{37}
\end{eqnarray}
where we have made use of the factorizability of Eq.\ (\ref{31}) and the
result of integration over $p_2$ given in Eq.\ (\ref{33}).  This is the
constant pedestal in $\Delta \eta$, which we relate to $\Delta T$.  Using
$C=23.2$ (GeV/c)$^{-1}$, $T_0 = 0.317$ GeV/c from \cite{hy},
$p_a = 2,\ p_b = 4$   GeV/c as in the experiment \cite{jaw}, and Eq.\
(\ref{21}) for the definition of $\Delta T$, we fit the pedestal height of
$\sim 0.05$ \cite{jaw} and obtain
\begin{eqnarray}
\Delta T = 15\ {\rm MeV} \quad.
\label{38}
\end{eqnarray}
Thus the inverse slope of the enhanced thermal partons is only 5\%
   higher than that of the un-enhanced partons, small enough to have a
negligible effect on the calculation of the single-particle distribution
using $T_0$.  However, the effect on the associated particle
distribution is evidently not negligible.  Although Eq.\ (\ref{38}) is a
fitted result, it should be emphasized that it is in the specific model of
parton recombination that the pedestal phenomenon is interpreted as a
consequence of the enhancement of the thermal energy of  the soft
partons due to energy loss of the hard parton.

Equation (\ref{37}) is the projection of $dN^{AP}/d\Delta \eta d \Delta
\phi$ onto $\Delta \eta$.  We now project it onto $\Delta \phi$ by
integrating it over $\Delta \eta$ from $-1$ to $+1$ in accordance to the
experimental cut \cite{jaw} and obtain from Eq.\ (\ref{31}) for the
pedestal part
\begin{eqnarray}
{dN^{ped}  \over d\Delta\phi } = \int^{1}_{-1}d\Delta\eta {dN^{ped}
\over d\Delta\eta d\Delta\phi } = { C^2 \over  3} \left[T^2 h\left({p_a  \over
T } , {p_b  \over T } \right)H(\Delta\phi)- T^2_0 h\left({p_a  \over
T_0}, {p_b  \over  T_0 } \right)  \right] \quad .
\label{39}
\end{eqnarray}
This is not a flat pedestal as in the case of $\Delta \eta$, but is
nevertheless a remnant of the effect of enhanced thermal partons sitting
under a peak from the $({\cal TS} + {\cal SS})_{24}$ part of $F^{AP}_4$
that we have not yet calculated.  Although it is not visible as a plateau,
its origin is the same as that which gives rise to the pedestal in $\Delta
\eta$.  It is the background subtraction in $\Delta \phi$ that forces
Eq.\ (\ref{39}) to vanish at $|\Delta\phi| = 1$, which is achieved by our
choice of $\sigma_{\phi}$ in Eq.\ (\ref{35}) its value is
\begin{eqnarray}
\sigma_{\phi} = 1.2 \ .
\label{40}
\end{eqnarray}
With this choice of $\sigma_{\phi}$ Eq.\ (\ref{39}) is entirely specified
numerically.  The $\Delta\phi$ dependence of
$dN^{ped}/d\Delta\phi$ is a broad mount between $\Delta\phi =
\pm 1$, which we shall exhibit later together with the complete
$dN/d\Delta\phi$.
This mount is the projection of a ridge in the 3D display of APD in 
$(\Delta\eta, \Delta\phi)$ onto the $\Delta\phi$ subspace.

This completes our discussion of the $({\cal TT}-{\cal T}_0 {\cal T}_0)$
contribution to $F^{AP}_4$ in Eq.\ (\ref{26}) and then to Eqs.\
(\ref{28}) and  (\ref{29}).  We now proceed to the last two terms of
$F^{AP}_4$.

\section{Peaks in $\Delta \eta$ and $\Delta \phi$}

We now return to Eq.\ (\ref{26}) and consider the $({\cal TS} + {\cal
SS})_{24}$ terms in $F^{AP}_4$ that we have put aside.  The ${\cal TS}$
component has already been described in detail in Eq.\ (\ref{20}).  The
${\cal SS}$ component is less important except in peripheral collisions.
It can be included by the replacement of the $\left\{S^j_i,
S^{j^{\prime}}_i \right\}{\cal T}$ term in Eq.\ (\ref{20}) by a $\left\{S^j_i,
S^{j^{\prime}}_i, S^{j^{\prime\prime}}_i \right\}$ term with appropriate
symmetrization. However, since the recombination of two shower partons
reproduces the fragmentation function in accordance to
\begin{eqnarray}
xD_i^{\pi}(x)=\int{dx_1\over x_1}{dx_2\over
x_2}\left\{S_i^j(x_1),S_i^{j'}\left({x_2\over
1-x_1}\right)\right\}R_{\pi}(x_1,x_2,x)\ ,  \label{41}
\end{eqnarray}
from which the SPDs are derived in the first place, we may replace a
$SSR$ term by a $xD$ term.
For notational simplicity we leave the $({\cal SS})_{24}$ terms out in
our description below, but numerically include their recombination in the
calculated result.

Let us recall that starting from the general formula Eq.\ (\ref{10}) we
first focussed on the variables in the plane containing $\vec{k}$ and the
longitudinal direction $\hat{z}$ and related the relevant angles to
pseudorapidities in Eq.\ (\ref{12}).  We then defined the angle $\psi$
between the shower parton $\vec{q}_2$ and the hard parton $\vec{k}$
to be as given in Eq.\ (\ref{14}).    The angular dependence of the SPD
is then expressed by a Gaussian distribution $G(\psi,x)$ in Eq.\
(\ref{17}), put in a factorized form in Eq.\ (\ref{16}).
We may identify $G(\psi,x)$ as the $x$-$z$ projection of a general
Gaussian distribution that describes the dependence on the angle between
  $\vec{p}_1$ and $\vec{p}_2$ in 3D geometry. In the small width
approximation, it can be shown that the corresponding projection onto the
transverse, $x$-$y$, plane, gives a Gaussian form, $G(\Delta\phi,x)$, for
the $\Delta\phi$ dependence.

Returning to Eq.\ (\ref{20}) which is written for the plane containing
$\vec{k}$ and $\hat{z}$, we now write the double differential in both
$\Delta \eta$ and $\Delta \phi$, as for the pedestal term, but now for the
peak term arising from the $({\cal TS})_{13}({\cal TS})_{24}$
component
\begin{eqnarray}
{dN^{peak}_{\pi^+ \pi^-}  \over  dp_1 dp_2 d\eta_1
d\Delta
\eta d \Delta \phi}   &=& {1
\over (p_1 p_2)^2}
\xi \sum_i \int dk k f_i(k) \int dq_1\int dq_2 \theta(k-q_1-q_2)
\nonumber\\ && \sum _{jj^{\prime}}  {\cal T} (p_1-q_1) \left\{ S^j_i
\left({q_1  \over k }
\right),  S^{j^{\prime}}_i \left({q_2  \over k-q_1 }\right)\right\}
{\cal T}(p_2
-q_2) \nonumber\\
&&\left.G\left(\psi , {q_2 \over k}\right)\right|_{\psi =
\tan^{-1}g(\eta,\eta_1)}bG\left(\Delta \phi,  {q_2 \over k}\right)
\quad ,
\label{42}
\end{eqnarray}
where $b$ is a numerical normalization factor such that when Eq.\
(\ref{42}) is integrated over $\Delta \phi$ from $-0.5$ to $+0.5$ (the
experimental window for projection to the $\Delta \eta$ dependence)
one gets
\begin{eqnarray}
b\int^{0.5}_{-0.5} d \Delta \phi  G(\Delta \phi,x = 0) = 1
\label{43}
\end{eqnarray}
in the small $x$ approximation.

Identifying Eq.\ (\ref{42}) as $N(1, 2)^{peak}$, we can substitute it into
the numerator of Eq.\ (\ref{29}) and obtain $dN^{peak}/d\Delta \eta d
\Delta \phi$.  There is only one free parameters to adjust; it is
$\sigma_0$ in Eq.\ (\ref{17}).  The projections of the double
differential distribution to $\Delta \eta$ and $\Delta \phi$ separately
can be compared with the data when the pedestal component is added.
That is
\begin{eqnarray}
{dN^{AP}  \over d\Delta\eta }= \int^{0,5}_{-0.5} d \Delta \phi
\left[{dN^{ped}
\over d\Delta\eta d\Delta\phi} + {dN^{peak}\over
d\Delta\eta d\Delta\phi}\right]
\quad ,
\label{44}
\end{eqnarray}
\begin{eqnarray}
{dN^{AP}  \over d\Delta\phi }= \int^{1}_{-1} d \Delta \eta
\left[{dN^{ped}
\over d\Delta\eta d\Delta\phi} + {dN^{peak}\over
d\Delta\eta d\Delta\phi}\right]
\quad.
\label{45}
\end{eqnarray}
The pedestal contributions to the above integrals are given by
Eqs.\ (\ref{37}) and  (\ref{39}), respectively.  It should be noted that the
latter has a $\Delta\phi$ dependence described by $H(\Delta \phi)$,
which is a Gaussian with a half-width $\sigma_{\phi}$ given by Eq.\
(\ref{40}).  We shall find below that the half-width of the peak term,
controlled by $\sigma_0$, is much smaller than $\sigma_{\phi}$, so the
last terms in Eqs.\ (\ref{44}) and  (\ref{45}) dominate the peak
structure of the APD.  The term $dN^{ped}/d\Delta\eta d\Delta\phi$
gives rise to
a ridge, whose projections are
the flat pedestal in $dN^{AP}/ d\Delta\eta$,
and a broad
mount in $dN^{AP}/ d\Delta\phi$.

After we put all the pieces together, using Eq.\ (\ref{29}) as the
important link between our calculation and the measurable quantities,
the results from  Eqs.\ (\ref{44}) and  (\ref{45}) can be compared to
data with $\sigma _0$ adjusted to fit the peak width.  Figure 1 shows
the APD in $\Delta \eta$.  The solid line is the result of our calculation
when we set
\begin{eqnarray}
\sigma _0 = 0.22\ .
\label{45a}
\end{eqnarray}
Evidently, it reproduces the data \cite{jaw} satisfactorily, both in the
peak structure and the pedestal.  It should be emphasized that although the
width and the pedestal height are adjusted to fit by our choice of $\sigma
_0$ and
$\Delta T$, the height of the peak is a consequence of our very
complicated calculation involving multiple integrals over many terms
based on the recombination model.  It is by no means trivial that the
data can be fitted so well.  In point of fact, we have only calculated
$\pi^+\pi^-$ production, whereas the data are for all charged particles.
Thus the precise values of the parameters are not as significant as the
overall situation where the pedestal and peak structure of the $\Delta
\eta$ distribution of the data can be reproduced by our description of
dihadron correlation.

Turning now to the APD in $\Delta \phi$, we have no free parameters to
adjust, since the double differentials in the square brackets in Eqs.\
(\ref{44}) and  (\ref{45}) are identical.  Upon integration over $\Delta
\eta$ we obtain the solid line in Fig.\ 2, which compares well with the
data \cite{jaw}.  The dashed line indicates the pedestal contribution
from Eq.\ (\ref{39}).  It plays the role of the flat pedestal in Fig.\ 1, but
here it vanishes at $|\Delta \phi| = 1$ because of the subtraction
scheme.  Since the peak structure in  $\Delta \phi$ is much narrower
than the broad mount of the pedestal (due to $\sigma _0 \ll \sigma
_{\phi}$), we have applied the vanishing condition at $|\Delta \phi| = 1$
only to the $dN^{ped}/ d\Delta\phi$ component, knowing that the
$dN^{peak}/ d\Delta\phi$ component is negligible at the wings of the
peak.

Figures 1 and 2 represent our main results on dihadron correlation in
central Au+Au collisions at $\sqrt{s} = 200$ GeV, when the trigger
particle is kept within the range $4 < p_{trig} < 6$ GeV/c
   and the associated particle in the range $2< p_{assoc}< 4$ GeV/c.  We
can readily calculate the APD at higher momentum ranges, but not at
lower momenta, since our model is not reliable for parton momentum
less than 1 GeV/c.  For that reason we do not calculate the APD for
$0.15 < p_{assoc} < 4$ GeV/c, even though data for that range are
presented in \cite{jaw}.

\section{Conclusion}
To summarize, we have successfully reproduced the data that show the
existence of peaks in $\Delta\eta$ and $\Delta\phi$ distributions,
the former sitting above a
flat pedestal, while the latter sitting above a broad mount.
  Some parameters are used to fit
the data, but the essence of our work is not data fitting. We have
demonstrated that the physics underlying the detail structure of the
jet characteristics observed in RHIC experiments can be understood in
the framework of parton recombination. Jets produced in heavy-ion
collisions create shower partons that are in the environment of
thermal partons, which can themselves be enhanced by the passage of
hard partons through the medium. That view is probably shared by all
theoretical approaches  to the problem. The issue then is how those
partons hadronize.  Different models treat the hadronization process
differently. Our approach emphasizes the recombination of the thermal
partons and the shower partons in the intermediate $p_T$ region. That
is our way of accounting for the medium effects on jets. So far we
have not encountered any obstacle in understanding the data in that
way. Data fitting is only a concrete demonstration that the details
of the jet structure can be quantitatively reproduced in our
treatment. The hadronization formalism is thereby enriched by the
determination of some features in the model by phenomenology.

Our first discovery in this work is that the pedestal in the
$\Delta\eta$ distribution can be related to $\Delta T$ in the local thermal
distribution. No ``long-range correlation" has been put in by hand.
The pedestal is a consequence of the recombination of the thermal
partons among themselves, which are only indirectly affected by hard
scattering through the enhancement of $T$ due to energy loss. Thus
the pedestal is not a part of the jet, but cannot be present without
a jet.
  This chain of successive connections involving energy loss, enhanced
thermal partons, their hadronization by recombination, and the elevated
APD in $\Delta\eta$ is certainly very different from a model where an
explicit correlation is put in by hand.

Our second achievement in this work is the success in describing the
peak structures in both $\Delta\eta$ and $\Delta\phi$ distributions
in terms of one angular distribution of the shower partons in a jet
cone. No ``short-range correlation" has been put in by hand. The only
primitive correlation in the problem is that which exists among the
shower partons in a jet.
  The properties of that correlation in the momentum magnitudes of the
shower partons have already been described in \cite{ht}. Here we show how
the correlations manifest themselves in the angular variables of the
produced pions.

There is one caveat in our analysis that should be noted. We have
calculated only $\pi^+\pi^-$ production, whereas the data are on all
charged hadrons. Thus numerically the parameters determined here are
not definitive. Since particle identification is steadily being
improved, the more appropriate arena for detailed matching of theory
and experiment is when the data for the production of specific
species become available.

Despite the tentativeness of the value of $\Delta T$ given in Eq.\
(\ref{38}) on account of the statement made just above, our assertion
that the pedestal is related to the enhancement of $T$ of the thermal
partons remains unaffected. As a test of that assertion we propose
that the proton associated with a $\pi^+$ trigger be measured in
addition to the measurement of a $\pi^+$ or $\pi^-$ associated
particle. The $p/\pi$ ratio of the associated particles in a $\pi^+$
triggered jet should be a good probe of the physical mechanism
underlying both the pedestal and the peak. The recombination approach
to the problem is well positioned to calculate that ratio, as it has
for single-particle distribution \cite{hy}. We surmise that the pedestal part
of the $p/\pi$ ratio would be higher than in the single-particle case
because $\Delta T>0$.

What we have done in this paper is influenced greatly by our
intention to understand the pedestal and peak structure in the data
of [3]. As a consequence, our analysis involves many integrals that
correspond to the experimental cuts in the data. Having determined
the origin of the pedestal and peaks in the  $\Delta\eta$ and
$\Delta\phi$ distributions, we are now equipped to launch a study of
the correlation problem independent of any triggers and related
experimental cuts, on which more and more data are becoming available.

\section*{Acknowledgment}
We have benefitted from discussions with X.\ N.\ Wang and C.\ B.\
Yang in the preliminary stage of this work. Communication with F.\
Wang has also been helpful.
    This work was supported, in part,  by the U.\ S.\ Department of Energy under
Grant No. DE-FG03-96ER40972.

\vskip1cm
\begin{center}
\section*{Figure Captions}
\end{center}

\begin{description}
\item
Fig.\ 1. Associated particle distribution in $\Delta\eta$ for 
$2<p_T<4$ GeV/c  with  trigger particle in $4<p_T^{trig}<6$ GeV/c. 
The data from Ref.\ \cite{jaw} are for all charged hadrons in the 
respective $p_T$ ranges. The solid line is the result of our 
calculation of $\pi^-$ associated with $\pi^+$ trigger.

\item
Fig.\ 2. Same as in Fig.\ 1 except that the distribution is in $\Delta\phi$.
The dashed line represents the pedestal effect in $\Delta\phi$ forced 
to vanish at $|\Delta\phi|=1$ by the subtraction scheme discussed in Sec.\ 3.

\end{description}

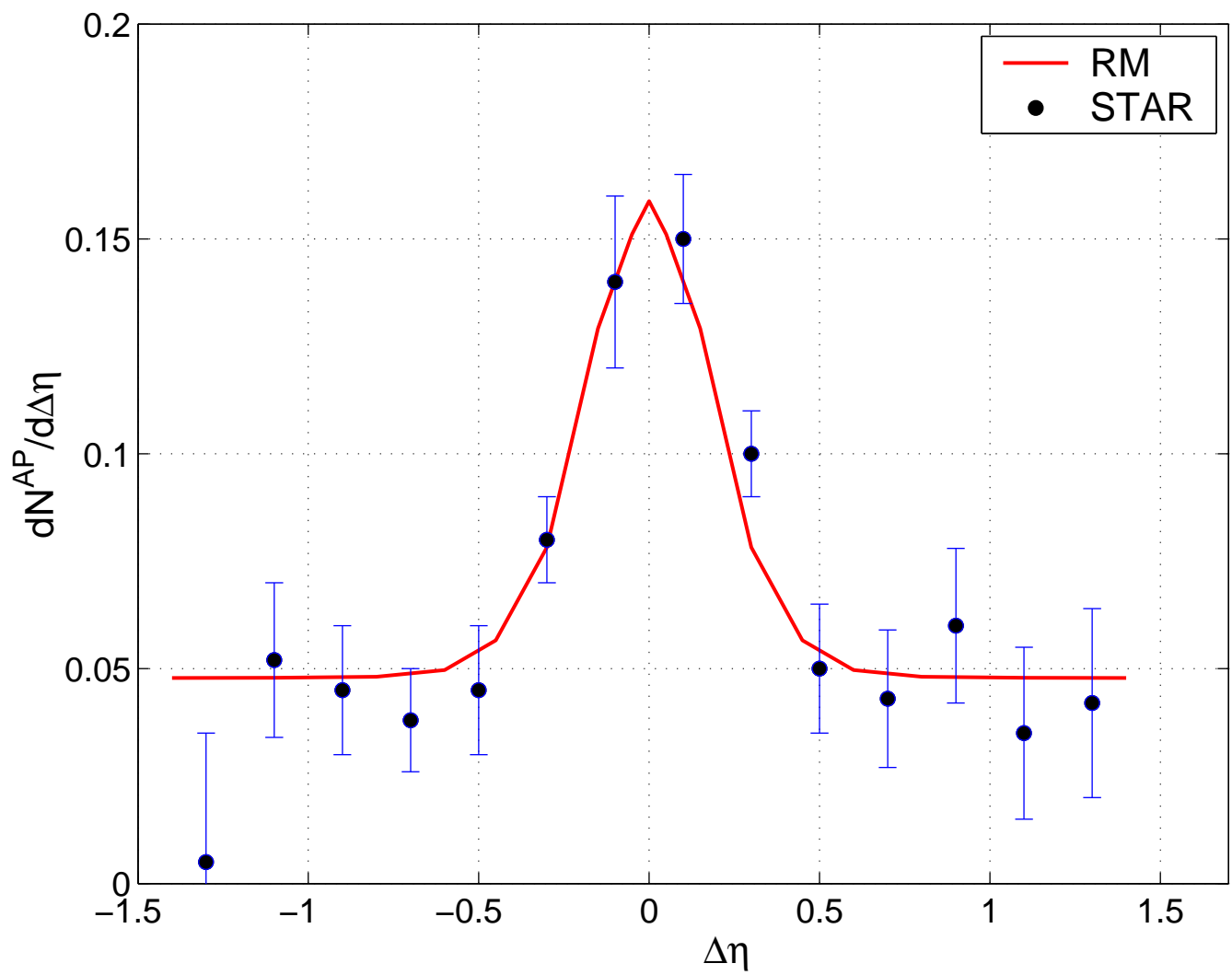
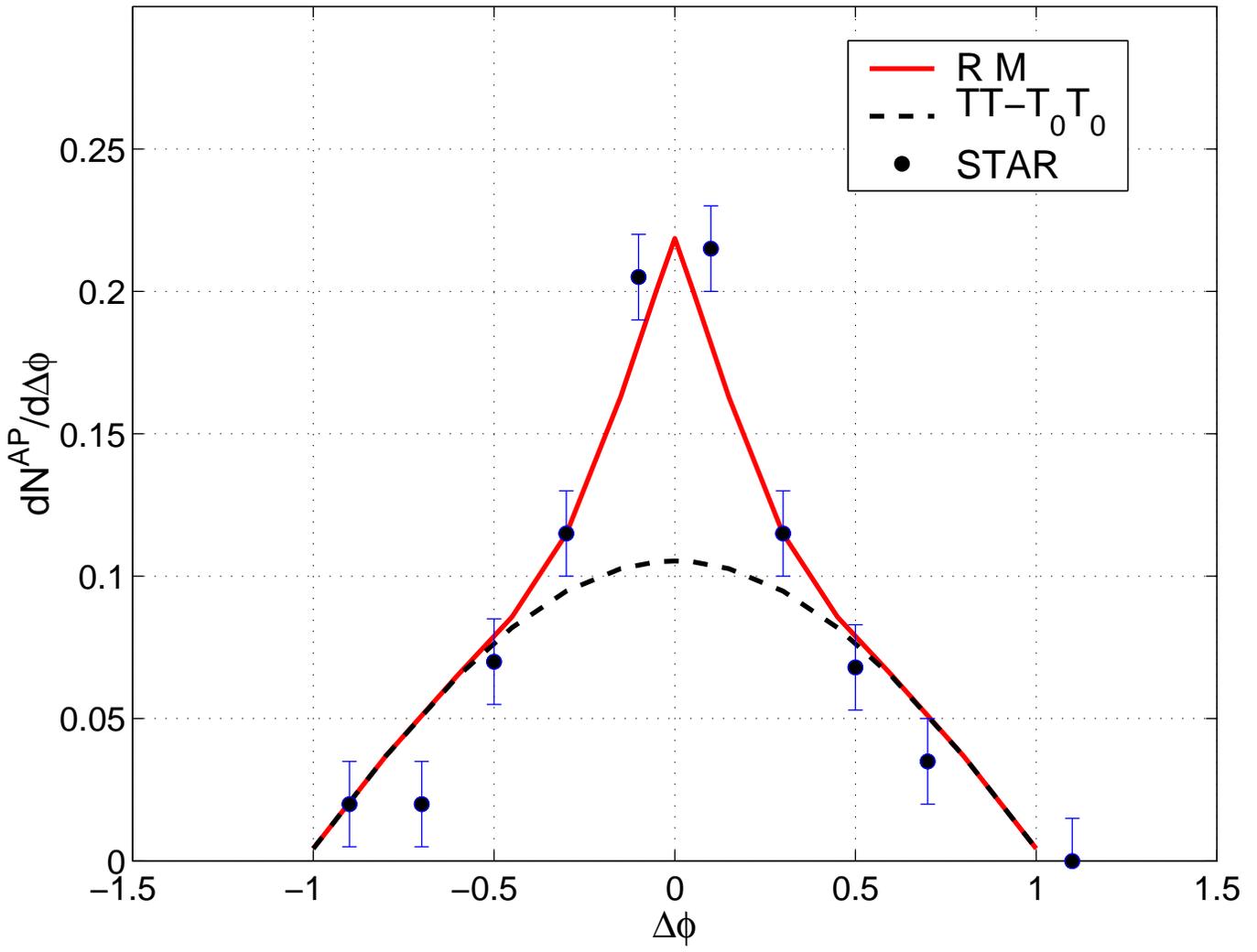
\end{document}